\documentclass[10pt,oneside]{article}
\usepackage{amsfonts,amssymb,graphicx}
\usepackage{cite}
\usepackage{indentfirst}
\usepackage{newlfont}
\usepackage{amsthm}
\usepackage{amsmath}
\usepackage{latexsym}
\usepackage{eucal}
\usepackage{eufrak}
\newtheorem{theorem}{Theorem}

\newtheorem{lemma}{Lemma}

\begin{document}

\begin{center}
\vspace{1truecm}
{\bf\Large Scaling Limits for Multi-Species Statistical Mechanics Mean-Field Models}\\
\vspace{1cm}
{Micaela Fedele and Pierluigi Contucci}\\
\vspace{.5cm}
{\small Dipartimento di Matematica} \\
{\small Universit\`a di Bologna, 40127 Bologna, Italy}\\
{\small {e-mail: {\em fedele@dm.unibo.it}, {\em contucci@dm.unibo.it}}}\\
\vskip 1truecm
\end{center}
\vskip 1truecm
\begin{abstract}\noindent
We study the limiting thermodynamic behavior of the normalized sums of spins
in multi-species Curie-Weiss models. We find sufficient conditions for
the limiting random vector to be Gaussian (or to have an exponential distribution of 
higher order) and compute the covariance matrix in terms of model parameters.\\

\end{abstract}

\section*{Introduction}
The study of the normalized sum of random variables and its asymptotic behavior has been
and continues to be a central chapter in probability and statistical mechanics. When those 
variables are independent and have finite variance the central limit theorem ensures that the sum with square-root normalization
converges toward a Gaussian distribution. The generalization of that result to non-independent
variables is particularly interesting in statistical mechanics where the random variables
have an Hamiltonian interaction.

Ellis and Newman \cite{ellis1978limit,ellis1978statistics,ellis1980limit} have studied the distribution 
of the normalized sums of spins whose interaction is described by a wide class of mean field Hamiltonian 
a la Curie-Weiss. They have found the conditions, in terms of the interaction, that lead in the thermodynamic limit 
to a Gaussian behavior and those which lead to a higher order exponential probability distribution. 

In recent times a multi-species extension of the Curie-Weiss model has been proposed
in the attempt to describe the large scale behavior of some socio-economic systems \cite{contucci2007modeling}. 
Multi-populated non-interacting spin models are at the basis of the so called Mc Fadden discrete choice
\cite{mcfadden2001economic} theory. The extension of the discrete choice theory to the interacting,
and more realistic, case is an important problem toward the understanding of the collective behavior
of social and economical systems. The investigation of multi-species models
has been pursued at a mathematical level in \cite{gallo2008bipartite} where there have been proved
properties like the existence of the thermodynamic limit by monotonicity, the computation of the free energy
and of the intensive quantities like local magnetizations. The phenomenological test of the model has been
started in \cite{gallo2008parameter} and it is a topic of current investigations.

In this paper we deal with the study of the normalized sum behavior for a multi-populated model
with mean field interaction. We prove that, under the assumption that the mean field Hamiltonian interaction
has a convexity property, when the system reaches its thermodynamic limit
the random vector whose components are the sums of spins on each population, converges 
to a nontrivial random variable $S$. The behavior of $S$ depends crucially upon the nature of the 
minima points of a function $G$ (the pressure functional) which we associate to the model interaction type. 
In particular it is the value of the determinant of the Hessian matrix of $G$ computed on the minima points
that establishes the Gaussian or non Gaussian behavior of the random vector. In the case of a unique
minimum point if the determinant is different from zero then $S$ is a multivariate Gaussian whose covariance
that can be computed from the mean field equations. Otherwise $S$ has a distribution whose density behaves as 
a higher order exponential. When the function $G$ has more than one minimum point we obtain a similar
result whenever the random vector of the magnetizations is close enough to
one of the minimum points. 

This work is organized as follows. Chapter one introduces the language and the notations
and states the main results in theorems \ref{teo1} and \ref{teo2}. Chapter 2 contains the proofs. Chapter
3 describes specific cases in which the distribution is Gaussian and others in which is not. 
The appendix contains the proof of the lemmas that make the paper self contained.

\section{Definitions and Statements}
We consider a system of $N$ particles that can be divided into $n$ subsets $P_{1},\dots , P_{n}$ with $P_{l}\cap P_{s}=\emptyset$, for $l\neq s$ and sizes $|P_{l}|=N_{l}$, where $\sum_{l=1}^{n}N_{l}=N$. Particles interact 
with each other and with an external field according to the mean field Hamiltonian:
\begin{equation}\label{hamiltoniana}
H_{N}(\boldsymbol{\sigma})=-\frac{1}{2N}\sum_{i,j=1}^{N}J_{ij}\sigma_{i}\sigma_{j}-\sum_{i=1}^{N}h_{i}\sigma_{i} \; .
\end{equation}
The $\sigma_{i}$ represents the spin of the particle $i$, $\sigma_{i}=\pm 1$ while $J_{ij}$ is the parameter that governs the mutual 
interaction between the particle $i$ and the particle $j$ and takes values according to the following symmetric matrix:

\begin{displaymath}
         \begin{array}{ll}
                \\
                N_1 \left\{ \begin{array}{ll||}
                                      \\
                                   \end{array}  \right.
                                        \\
                N_2 \left\{ \begin{array}{ll||}
                                        \\
                                   \end{array}  \right.
                                          \\
                                         \\
                                         \\
                      
                  N_n \left\{ \begin{array}{ll||}
                     \\
            \\
                \\
                                  \end{array}  \right.
         \end{array}
          \!\!\!\!\!\!\!\!
         \begin{array}{ll||}
                \quad
                 \overbrace{\qquad }^{\textrm{$N_1$}}\;
                 \overbrace{\qquad }^{\textrm{$N_2$}}\qquad\quad\;\;\;\;
                 \overbrace{\qquad\qquad\quad }^{\textrm{$N_n$}}
                  \\
                 \left(\begin{array}{c|c|cc|ccc}
                               \mathbf{ J}_{11}  &  \mathbf{ J}_{12} & &\;\dots\; & &\;\;\mathbf{ J}_{1n}\;\;&
                                \\
                                 \hline
                              \mathbf{ J}_{12}^{t} & \mathbf{ J}_{22} & & & & &\\
                             \hline
                             & & & & & &\\
                             \vdots & & & & & &\\
                             \hline
                             & & & & & &\\
                             \mathbf{ J}_{1n}^{t} & \mathbf{ J}_{2n}^{t} & &\;\dots\; & &\;\;\mathbf{ J}_{nn}\;\; &\\
                             & & & & & &
                      \end{array}\right)
               \end{array}
\end{displaymath}\\
\noindent where each block $\mathbf{J}_{ls}$ has constant elements $J_{ls}$. For $l=s$, $\mathbf{J}_{ll}$ is a square matrix, whereas the matrix $\mathbf{ J}_{ls}$ is rectangular. We assume $J_{11}, J_{22},\dots , J_{nn}$ be positive, whereas $J_{ls}$ with $l\neq s$ can be either positive or negative allowing both ferromagnetic and antiferromagnetic interactions. The vector field takes also different values depending on the subset the particles belong to: 

\begin{displaymath}
         \begin{array}{ll}
                N_1 \left\{ \begin{array}{ll}
                                      \\
                                   \end{array}  \right.
                                        \\
                N_2 \left\{ \begin{array}{ll}
                                        \\
                                   \end{array}  \right.
                                          \\
                                         \\
                                         \\
                      
                  N_n \left\{ \begin{array}{ll}
                     \\
            \\
                \\
                                  \end{array}  \right.

           \!\!\!\!\!\!
    \end{array}
    \!\!\!\!\!\!
    \left(\begin{array}{ccc|c}
                \mathbf{h}_{1}
            \\
            \hline
            
            \mathbf{h}_{2}
            \\
            \hline
            \\
            \vdots
            \\
            \hline
            \\
            \mathbf{h}_{n}
            \\
            \\
        \end{array}\right)
\end{displaymath}\\
\noindent where each $\mathbf{h}_{l}$ is a vector of constant elements $h_{l}$.

The joint distribution of a spin configuration $\boldsymbol{\sigma}=(\sigma_{1},\dots ,\sigma_{N})$ is given by the Boltzmann-Gibbs measure:
\begin{equation}\label{distribuzionespin}
P_{N,\mathbf{J},\mathbf{h}}\{\boldsymbol{\sigma}\}=\frac{1}{Z_{N}(\mathbf{J},\mathbf{h})}e^{-H_{N}(\boldsymbol{\sigma})}\prod\limits_{i=1}^{N}d\rho(\sigma_{i})
\end{equation}
where $Z_{N}(\mathbf{J},\mathbf{h})$ is the partition function
\begin{equation}\label{funzpartizione}
Z_{N}(\mathbf{J},\mathbf{h})=\int\limits_{\mathbb{R}^{N}}e^{-H_{N}(\boldsymbol{\sigma})}\prod\limits_{i=1}^{N}d\rho(\sigma_{i})
\end{equation}
and $\rho$ is the measure:
\begin{equation}\label{misuraro}
\rho(x)=\frac{1}{2}\Big(\delta(x-1)+\delta(x+1)\Big)
\end{equation}
where $\delta(x-x_{0})\;\;x_{0}\in\mathbb{R}$ denotes the unit point mass with support at $x_{0}$. The inverse temperature $\beta$ isn't explicitly written because it is included in the parameters of the Hamiltonian.

By introducing the magnetization of a set of spins $A$ as:
\begin{equation}
m_{A}(\boldsymbol{\sigma})=\frac{1}{|A|}\sum_{i \in A}\sigma_{i}
\end{equation}
\noindent and indicating by $m_{l}(\boldsymbol{\sigma})$ the magnetization of the set $P_{l}$, and by $\alpha_{l}=N_{l}/N$ the relative size of the set $P_{l}$, we may easily express the Hamiltonian (\ref{hamiltoniana}) as:
\begin{equation}\label{Hamiltoniana2}
 H_{N}(\boldsymbol{\sigma})=-Ng\Big(m_{1}(\boldsymbol{\sigma}),\dots,m_{n}(\boldsymbol{\sigma})\Big)
\end{equation}
\noindent where the function $g$ is:
\begin{equation}\label{funzioneg}
g\Big(m_{1}(\boldsymbol{\sigma}),\dots,m_{n}(\boldsymbol{\sigma})\Big)=\frac{1}{2}\sum\limits_{l, s=1}^{n}\alpha_{l}\alpha_{s}J_{ls}m_{l}(\boldsymbol{\sigma})m_{s}(\boldsymbol{\sigma})+\sum\limits_{l=1}^{n}\alpha_{l}h_{l}m_{l}(\boldsymbol{\sigma}) \; .
\end{equation}

In \cite{gallo2008bipartite} it is shown that the thermodynamic limit of the pressure function
\begin{equation}
p_{N}(\mathbf{J},\mathbf{h})=\frac{1}{N}\ln \Big(\sum_{\boldsymbol{\sigma}}e^{-H_{N}(\boldsymbol{\sigma})}\Big)
\end{equation}
\noindent exists and is reached monotonically if the the function $g$ given by (\ref{funzioneg}) is convex (see also \cite{bianchi2003thermodynamic}). In this case: 
\begin{equation}
 \lim_{N\rightarrow\infty}p_{N}(\mathbf{J},\mathbf{h})=\sup_{\mathbf{x}\in [-1,1]^{n}}\overline{p}(\mathbf{x})
\end{equation}
\noindent where the functional $\overline{p}$ is:
\begin{equation}
\overline{p}(x_{1},\dots,x_{n}) =\ln
2-\frac{1}{2}\sum_{l, s=1}^{n}\alpha_{l}\alpha_{s}J_{ls}x_{l}x_{s}+\sum_{l=1}^{n}\alpha_{l}\ln\bigg(\cosh\bigg(\sum_{s=1}^{n}\alpha_{s}J_{ls}x_{s}+h_{l}\bigg)\bigg) \; .
\end{equation}

It is easy to verify that the function $g$ is convex if the following matrix, called reduced interaction matrix, 

\begin{equation}\label{interazioneridotta}
 \mathbf{J}=\begin{pmatrix}
J_{11}  & J_{12} & \dots & J_{1n}\\
J_{12}  & J_{22} & \dots & J_{2n}\\
\vdots&\vdots&&\vdots \\
J_{1n}  & J_{2n} & \dots & J_{nn}
\end{pmatrix}
\end{equation}\\
\noindent is positive definite.

The extremality conditions of $\overline{p}$ give the Mean Field Equations of the model
\begin{equation}\label{campomedio}
\begin{cases}
\mu_{1} &\!\!\!\!= \tanh\Big(\sum\limits_{l=1}^{n}\;\alpha_{l}J_{1l}\;\mu_{l}+h_{1}\Big) \\
\mu_{2} &\!\!\!\!=\tanh\Big(\sum\limits_{l=1}^{n}\;\alpha_{l}J_{2l}\;\mu_{l}+h_{2}\Big)\\
\;\vdots\\
\mu_{n} &\!\!\!\!=\tanh\Big(\sum\limits_{l=1}^{n}\;\alpha_{l}J_{ln}\;\mu_{l}+h_{n}\Big) \; .
\end{cases} 
\end{equation}

In the thermodynamic limit the random vector $(m_{1}(\boldsymbol{\sigma}),\dots,m_{n}(\boldsymbol{\sigma}))$ weakly converges, with respect to the Boltzmann-Gibbs measure, to the deterministic vector $(\mu_{1},\dots ,\mu_{n})$ solution of the Mean Field Equations. This means that the variances of the magnetizations vanish for large $N$ a part on isolated singularities (see \cite{gallo2008bipartite} for the precise statement). In this paper, define the sum of the spins of a set $A$ as:
\begin{equation}
S_{A}(\boldsymbol{\sigma})=\sum_{i \in A}\sigma_{i}
\end{equation}
\noindent and indicating by $S_{l}(\boldsymbol{\sigma})$ the sum of the spins of the set $P_{l}$ we want to determine a suitable normalization for the vector elements $S_{1}(\boldsymbol{\sigma}),\dots,S_{n}(\boldsymbol{\sigma})$ so that in the thermodynamic limit they converge to well defined random variables with finite (non zero) covariance matrix. 
The problem in $n=1$ has been solved in \cite{ellis1978limit} and \cite{ellis1980limit}.\\
We shall see that the behavior of the limiting distribution of the sums of spins depends crucially on the number and the nature of the minima points of the function $G=-\overline{p}+\ln2$  
\begin{equation}\label{funzioneG}
G(x_{1},\dots,x_{n})=\frac{1}{2}\sum_{l, s=1}^{n}\alpha_{l}\alpha_{s}J_{ls}x_{l}x_{s}-\sum_{l=1}^{n}\alpha_{l}\ln\bigg(\cosh\bigg(\sum_{s=1}^{n}\alpha_{s}J_{ls}x_{s}+h_{l}\bigg)\bigg)\text{.}
\end{equation}

Let $\boldsymbol{\mu}^{1},\dots,\boldsymbol{\mu}^{P}$ be global minima points of the function $G$. For each $p$ there exist the functions $G_{2j}^{\boldsymbol{\mu}^{p}}(\mathbf{x})\geq 0$, such that around $\boldsymbol{\mu}^{p}$ we can write $G$ as:
\begin{equation}\label{espansione.G.multi}
G(\mathbf{x}) =G(\boldsymbol{\mu}^{p})+\sum_{j=0}^{d}G_{2j}^{\boldsymbol{\mu}^{p}}(\mathbf{x}-\boldsymbol{\mu}^{p})+o\bigg(\Big(|\mathbf{x}'-\boldsymbol{\mu}^{p'}|^{2}+|\mathbf{x}''-\boldsymbol{\mu}^{p''}|^{2/q}\Big)^{d}\bigg)
\end{equation}
where $(\mathbf{x}',\mathbf{x}'')$ is a partition of the coordinate $\mathbf{x}$ and $q$ is a positive rational number such that $1/q\in\mathbb{N}$ and
\begin{equation}
 G_{2j}^{\boldsymbol{\mu}^{p}}(t\mathbf{x}',t^{q}\mathbf{x}'')=t^{2j}G_{2j}^{\boldsymbol{\mu}^{p}}(\mathbf{x}',\mathbf{x}'')\quad\quad\text{all}\;\;t>0.
\end{equation}
We define the type $k_{p}$ of the minimum point $\boldsymbol{\mu}^{p}$ as the smallest $d$ such that $G_{2d}^{\boldsymbol{\mu}^{p}}(\mathbf{x}-\boldsymbol{\mu}^{p})\neq 0$ as $\mathbf{x}\neq 0$ and $G_{2j}^{\boldsymbol{\mu}^{p}}(\mathbf{x}-\boldsymbol{\mu}^{p})=0$ for $j=1,\dots,d-1$.
We observe that when $q=1$ the expression (\ref{espansione.G.multi}) is the Taylor expansion of the function $G$. In this case, the 
only one we treat in this paper, $k_{p}$ is called the homogeneous type of the minimum point $\boldsymbol{\mu}^{p}$. In particular if a minimum points $\boldsymbol{\mu}^{p}$ has homogeneous type equal to $1$, around $\boldsymbol{\mu}$ we have:\\
\begin{equation}\label{espansione.G.multi.2}
G(\mathbf{x})=G(\boldsymbol{\mu}^{p})+\frac{1}{2}\langle\boldsymbol{\mathcal{H}}_{G}(\boldsymbol{\mu}^{p})(\mathbf{x}-\boldsymbol{\mu}^{p}),(\mathbf{x}-\boldsymbol{\mu}^{p})\rangle+o\Big(||(\mathbf{x}-\boldsymbol{\mu}^{p})^{2}||\Big)
\end{equation}\\
\noindent where $\boldsymbol{\mathcal{H}}_{G}(\boldsymbol{\mu}^{p})$ is the Hessian matrix of $G$ computed in the minimum point $\boldsymbol{\mu}^{p}$. 
 
We introduce some useful notations.
Considering $\mathbf{x},\mathbf{y}\in\mathbb{R}^{n}$ and $\gamma\in\mathbb{R}$ we define
\begin{itemize}
\item $\mathbf{x}^{\gamma}=(x_{1}^{\gamma},\dots,x_{n}^{\gamma})$;
\item $\mathbf{x}\mathbf{y}=(x_{1}y_{1},\dots,x_{n}y_{n})$;
\item $\dfrac{\mathbf{x}}{\mathbf{y}}=\Big(\dfrac{x_{1}}{y_{1}},\dots,\dfrac{x_{n}}{y_{n}}\Big)$ where $y_{l}\neq 0$ for $l=1,\dots,n$.
\end{itemize}

Now we can state our main results. The following theorem solves the problem of the correct normalization of the random vector of the sums of spins whenever the function $G$ admits a unique minimum point.
\begin{theorem}\label{teo1}
	Consider the mean-field Hamiltonian $H_{N}=-Ng(m_{1}(\boldsymbol{\sigma}),\dots,m_{n}(\boldsymbol{\sigma}))$ where $g$ is the convex function defined in (\ref{funzioneg}). Let $\boldsymbol{\mu}=(\mu_{1},\dots,\mu_{n})$ be the unique global minimum point of the function $G$ given by (\ref{funzioneG}). Let $k$ be the homogeneous type of the minimum point. 
	\begin{enumerate}
	\item If $k=1$ the random vector 
	\begin{equation}\label{vettore.S.1}
	\bar{\mathbf{S}}^{1}(\boldsymbol{\sigma})=\bigg(\dfrac{S_{1}(\boldsymbol{\sigma})-N_{1}\mu_{1}}{\sqrt{N_{1}}}, \dots,\dfrac{S_{n}(\boldsymbol{\sigma})-N_{n}\mu_{n}}{\sqrt{N_{n}}}\bigg)
	\end{equation}
	as $N_{1}\rightarrow\infty$, $\dots$, $N_{n}\rightarrow\infty$, for fixed values of $\alpha_{1},\dots, \alpha_{n}$, weakly converges to a normal multivariate distribution whose covariance matrix is:\\
	\begin{equation}\label{covarianza}
	\widetilde{\boldsymbol{\chi}}=\begin{pmatrix}
	\dfrac{\partial\mu_{1}}{\partial h_{1}} & \sqrt{\dfrac{\partial \mu_{1}}{\partial h_{2}}\;\dfrac{\partial \mu_{2}}{\partial h_{1}}} & \dots & \sqrt{\dfrac{\partial \mu_{1}}{\partial h_{n}}\;\dfrac{\partial \mu_{n}}{\partial h_{1}}}\\\\
	\sqrt{\dfrac{\partial \mu_{1}}{\partial h_{2}}\;\dfrac{\partial \mu_{2}}{\partial h_{1}}} & \dfrac{\partial \mu_{2}}{\partial h_{2}} & \dots & \sqrt{\dfrac{\partial \mu_{2}}{\partial h_{n}}\;\dfrac{\partial \mu_{n}}{\partial h_{2}}}\\\\
	\vdots&\vdots&&\vdots \\\\
	\sqrt{\dfrac{\partial \mu_{1}}{\partial h_{n}}\;\dfrac{\partial \mu_{n}}{\partial h_{1}}} & \sqrt{\dfrac{\partial \mu_{2}}{\partial h_{n}}\;\dfrac{\partial \mu_{n}}{\partial h_{2}}} & \dots & \dfrac{\partial \mu_{n}}{\partial h_{n}}
	\end{pmatrix}\begin{matrix}
	\\\\\\\\\\\\\\.
	\end{matrix}
	\end{equation}
	\item If $k>1$ the random vector 
	\begin{equation}\label{vettore.S.k}
	\bar{\mathbf{S}}^{k}(\boldsymbol{\sigma})=\bigg(\dfrac{S_{1}(\boldsymbol{\sigma})-N_{1}\mu_{1}}{(N_{1})^{1-1/2 k}}, \dots,\dfrac{S_{n}(\boldsymbol{\sigma})-N_{n}\mu_{n}}{(N_{n})^{1-1/2 k}}\bigg)
	\end{equation}
	\noindent as $N_{1}\rightarrow\infty$, $\dots$, $N_{n}\rightarrow\infty$, for fixed values of $\alpha_{1}$, $\dots$, $\alpha_{n}$, weakly converges to a distribution proportional to:
	\begin{equation}\label{densita}
	\exp\bigg(-G_{2k}^{\boldsymbol{\mu}}\Big(\dfrac{\mathbf{x}}{\boldsymbol{\alpha}^{1/2k}}\Big)\bigg)
	\end{equation}
	\noindent where $\boldsymbol{\alpha}=(\alpha_{1},\dots ,\alpha_{n})$.
	\end{enumerate}
\end{theorem}

The following second theorem handles the case in which the function $G$ reaches the minimum in more than one point.
\begin{theorem}\label{teo2}
	Consider the mean-field Hamiltonian $H_{N}=-Ng(m_{1}(\boldsymbol{\sigma}),\dots,m_{n}(\boldsymbol{\sigma}))$ where $g$ is the convex function defined in (\ref{funzioneg}). Let $\boldsymbol{\mu}=(\mu_{1},\dots,\mu_{n})$ be a global minimum point of the function $G$ given from (\ref{funzioneG}). Let $k$ be the homogeneous type of the minimum point. Define $\bar{\delta}$ to be the minimum distance between all distinct pair of global minimum points of the function $G$. Then for any $d\in(0,\bar{\delta})$ when the random vector of the magnetizations $(m_{1}(\boldsymbol{\sigma}),\dots,m_{n}(\boldsymbol{\sigma}))$ is inside the ball $B(\boldsymbol{\mu},d)$ centered in $\boldsymbol{\mu}$ of radius $d$
	\begin{enumerate}
	\item  if $k=1$ the random vector $\bar{\mathbf{S}}^{1}(\boldsymbol{\sigma})$ defined in (\ref{vettore.S.1}) as $N_{1}\rightarrow\infty$, $\dots$, $N_{n}\rightarrow\infty$, for fixed values of $\alpha_{1}$, $\dots$, $\alpha_{n}$, weakly converges to a normal multivariate distribution whose covariance matrix is given by (\ref{covarianza});
	\item if $k>1$ the random vector $\bar{{S}}_{k}(\boldsymbol{\sigma})$ defined in (\ref{vettore.S.k}) as $N_{1}\rightarrow\infty$, $\dots$, $N_{n}\rightarrow\infty$, for fixed values of $\alpha_{1}$, $\dots$, $\alpha_{n}$, weakly converges to a distribution proportional to:
	\begin{equation}
	\exp\bigg(-G_{2k}^{\boldsymbol{\mu}}\Big(\dfrac{\mathbf{x}}{\boldsymbol{\alpha}^{1/2k}}\Big)\bigg)
	\end{equation}
	where $\boldsymbol{\alpha}=(\alpha_{1},\dots ,\alpha_{n})$.
	\end{enumerate}
\end{theorem}

\section{Proofs}

\subsection{Proof of Theorem \ref{teo1}}
To prove the theorem we need the following lemmas.

\begin{lemma}\label{lemma1}
Suppose that for each $N$, $\mathbf{x}_{N}=(X_{N}^{(1)},\dots, X_{N}^{(n)})$ and $\mathbf{y}_{N}=(Y_{N}^{(1)},\dots, Y_{N}^{(n)})$ are independent random vectors. Suppose that $\mathbf{x}_{N}$ weakly converges to a distribution $\nu$ such that
\begin{equation}
 \int\limits_{\mathbb{R}^{n}} e^{i\langle\mathbf{r},\mathbf{x}\rangle}d\nu(\mathbf{x})\neq 0 \quad\quad\text{for all}\;\;\mathbf{r}\in\mathbb{R}^{n}\;.
\end{equation}
Then $\mathbf{y}_{N}$ weakly converges to $\mu$ if and only if $\mathbf{x}_{N}+\mathbf{y}_{N}$ weakly converges to the convolution $\nu *\mu$ of the distributions $\nu$ and $\mu$.
\end{lemma}
This result follows because the weak convergence of measures is equivalent to pointwise convergence of characteristic functions (see \cite{durrett2010probability}).

\begin{lemma}\label{proposizione}
If the reduced interaction matrix $\mathbf{J}$ of a model defined by the Hamiltonian (\ref{Hamiltoniana2}) is positive definite, then the function $G$ given in (\ref{funzioneG}) has a finite number (different from zero) of global minimum points and for any $N\in\{1,2,\dots\}$
\begin{equation}\label{proprietaG}
\int\limits_{\mathbb{R}^{n}}\exp\Big(\!-NG(x_{1},\dots,x_{n})\Big)dx_{1}\dots dx_{n}<\infty
\end{equation}
\end{lemma}
See appendix for the proof.

\begin{lemma}\label{lemma2}
Let $\mathbf{A}=\mathbf{D}_{\boldsymbol{\alpha}}\mathbf{J}\mathbf{D}_{\boldsymbol{\alpha}}$ be a positive definite matrix where the matrix $\mathbf{D}_{\boldsymbol{\alpha}}=diag\{\sqrt{\alpha_{1}},\dots,\sqrt{\alpha_{n}}\}$ and the matrix $\mathbf{J}$ is defined in (\ref{interazioneridotta}). Given the random vector $(W_{1},\dots, W_{n})$ whose joint distribution is the normal multivariate
\begin{equation}\label{normalebivariata}
\rho(\mathbf{x})=\sqrt{\frac{\det \mathbf{A}}{(2\pi)^{n}}}\; \exp\bigg(\!-\frac{1}{2}\langle\mathbf{A}\mathbf{x},\mathbf{x}\rangle\bigg)
\end{equation}

\noindent if $(W_{1},\dots, W_{n})$ is independent of $(S_{1}(\boldsymbol{\sigma}),\dots, S_{n}(\boldsymbol{\sigma}))$ then for $(\mu_{1},\dots,\mu_{n}) \in \mathbb{R}^{n}$ and $\gamma\in \mathbb{R}$ the joint distribution of\\
\begin{equation}\label{sommavettori}
\Big(\dfrac{W_{1}}{(N_{1})^{1/2-\gamma}},\dots,\dfrac{W_{n}}{(N_{n})^{1/2-\gamma}}\bigg)+\bigg(\dfrac{S_{1}(\boldsymbol{\sigma})-N_{1}\mu_{1}}{(N_{1})^{1-\gamma}},\dots, \dfrac{S_{n}(\boldsymbol{\sigma})-N_{n}\mu_{n}}{(N_{n})^{1-\gamma}}\Big)
\end{equation}\\
\noindent is given by 
\begin{equation}\label{distribuzione.tesi}
\dfrac{\exp\bigg(\!-NG\Big(\dfrac{x_{1}}{N_{1}^{\;\gamma}}+\mu_{1},\dots,\dfrac{x_{n}}{N_{n}^{\;\gamma}}+\mu_{n}\Big)\bigg)dx_{1}\dots dx_{n}}{\displaystyle{\int\limits_{\mathbb{R}^{n}}} \exp\bigg(\!-NG\Big(\dfrac{x_{1}}{N_{1}^{\;\gamma}}+\mu_{1},\dots,\dfrac{x_{n}}{N_{n}^{\;\gamma}}+\mu_{n}\Big)\bigg)dx_{1}\dots dx_{n}}
\end{equation}\\
\noindent where $G$ is the function defined in (\ref{funzioneG}).
\end{lemma}
See appendix for the proof.

We remark that as $\gamma\!<\!1/2$, the random vector $(W_{1},\dots, W_{n})$ does not contribute to the limit of (\ref{distribuzione.tesi}) as $N_{1}\rightarrow\infty$, $\dots$, $N_{n}\rightarrow\infty$.

To prove the previous theorem we proceed as follows.
For $k>1$, by lemmas \ref{lemma1} and \ref{lemma2} with $\gamma=1/2k$, we have to prove that, for any bounded continuous function $\psi(\mathbf{x}):\mathbb{R}^{n}\rightarrow\mathbb{R}$

\begin{equation}\label{dadimostrare1}
	\dfrac{\displaystyle{\int_{\mathbb{R}^{n}}} \exp\bigg(\!-NG\Big(\dfrac{\mathbf{x}}{\mathbf{N}^{1/2k}}+\boldsymbol{\mu}\Big)\bigg)\psi(\mathbf{x})d\mathbf{x}}{\displaystyle{\int_{\mathbb{R}^{n}}} \exp\bigg(\!-NG\Big(\dfrac{\mathbf{x}}{\mathbf{N}^{1/2k}}+\boldsymbol{\mu}\Big)\bigg)d\mathbf{x}}
	\rightarrow \dfrac{\displaystyle{\int_{\mathbb{R}^{n}}} \exp\bigg(-G_{2k}^{\boldsymbol{\mu}}\bigg(\frac{\mathbf{x}}{\boldsymbol{\alpha}^{1/2k}}\bigg)\bigg)\psi(\mathbf{x})d\mathbf{x}}{\displaystyle{\int_{\mathbb{R}^{n}}} \exp\bigg(-G_{2k}^{\boldsymbol{\mu}}\bigg(\frac{\mathbf{x}}{\boldsymbol{\alpha}^{1/2k}}\bigg)\bigg)d\mathbf{x}}
	\end{equation}
where to easy the notation we set $\mathbf{N}=(N_{1},\dots,N_{n})$.
Defined the function:
\begin{equation}\label{funzione.B.multi}
B(\mathbf{x};\boldsymbol{\mu})=G(\mathbf{x}+\boldsymbol{\mu})-G(\boldsymbol{\mu})
\end{equation}

\noindent there exists $\delta>0$ sufficiently small so that, as $N_{1}\rightarrow\infty$, $\dots$, $N_{n}\rightarrow\infty$ for $||\mathbf{x}/\mathbf{N}^{1/2k}||<\delta$\\
\begin{align}\label{proprieta.B.multi}
N\cdot B\bigg(\dfrac{\mathbf{x}}{\mathbf{N}^{1/2k}},\boldsymbol{\mu}\bigg)&=G_{2k}^{\boldsymbol{\mu}}\bigg(\frac{\mathbf{x}}{\boldsymbol{\alpha}^{1/2k}}\bigg)+o(1)P_{2k}(\mathbf{x})\nonumber\\\\
N\cdot B\bigg(\dfrac{\mathbf{x}}{\mathbf{N}^{1/2k}},\boldsymbol{\mu}^{p}\bigg)&\geq\frac{1}{2}G_{2k}^{\boldsymbol{\mu}}\bigg(\frac{\mathbf{x}}{\boldsymbol{\alpha}^{1/2k}}\bigg)+o(1)P_{2k-1}(\mathbf{x})\nonumber
\end{align}
\noindent  where $P_{2k}(\mathbf{x})$ is a polynomial of $2k$ order and $P_{2k-1}(\mathbf{x})$ is a polynomial of $2k-1$ order.\\
Defined $f=G(\mathbf{\boldsymbol{\mu}})$, for any closed subset $V$ of $\mathbb{R}^{n}$ which does not contain $\mathbf{\boldsymbol{\mu}}$ there exist $\epsilon >0$, so that as $N\rightarrow\infty$
\begin{equation}\label{fuoriminimo}
e^{Nf}\int_{V}e^{-NG(\mathbf{x})}d\mathbf{x}=O\Big(e^{-N\epsilon}\Big)\;.
\end{equation}
\noindent We pick $\delta >0$ as in (\ref{proprieta.B.multi}). By (\ref{fuoriminimo}) there exists $\epsilon >0$ so that
\begin{equation}
e^{Nf}\int_{||\frac{\mathbf{x}}{\mathbf{N}^{1/2k}}||\geq\delta}\exp\bigg(\!-NG\Big(\dfrac{\mathbf{x}}{\mathbf{N}^{1/2k}}+\boldsymbol{\mu}\Big)\bigg)\psi(\mathbf{x})d\mathbf{x}
=O\Big(\Big(\prod_{l=1}^{n} N_{l}\Big)^{1/2k}e^{-N\epsilon}\Big)
\end{equation}
\noindent  whereas by (\ref{proprieta.B.multi}) and dominate convergence, we have that:
\begin{align*}
e^{Nf}\int_{||\frac{\mathbf{x}}{\mathbf{N}^{1/2k}}||<\delta} \!\!\!\exp\bigg(\!-NG\Big(\dfrac{\mathbf{x}}{\mathbf{N}^{1/2k}}+\boldsymbol{\mu}\Big)\bigg)\psi(\mathbf{x})d\mathbf{x}
&=e^{N(f-G(\boldsymbol{\mu}))}\int_{||\frac{\mathbf{x}}{\mathbf{N}^{1/2k}}||<\delta} \!\!\!\exp\bigg(\!-NB\Big(\dfrac{\mathbf{x}}{\mathbf{N}^{1/2k}},\boldsymbol{\mu}\Big)\bigg)\psi(\mathbf{x})d\mathbf{x}\\
&\rightarrow\int_{\mathbb{R}} \exp\bigg(\!-G^{\boldsymbol{\mu}}_{2k}\Big(\dfrac{\mathbf{x}}{\boldsymbol{\alpha}^{1/2k}}\Big)\bigg)\psi(\mathbf{x})d\mathbf{x}\;.
\end{align*}

\noindent This proves the statement $(2)$ of the theorem.

We observe that for $k=1$
\begin{equation*}
	 G_{2k}^{\boldsymbol{\mu}}\bigg(\frac{\mathbf{x}}{\boldsymbol{\alpha}^{1/2k}}\bigg)=\frac{1}{2}\langle\boldsymbol{\widetilde{\mathcal{H}}}_{f}(\boldsymbol{\mu})\mathbf{x},\mathbf{x}\rangle
	\end{equation*}
\noindent where $\boldsymbol{\widetilde{\mathcal{H}}}_{G}=\mathbf{D}_{\boldsymbol{\alpha}}^{-1}\boldsymbol{\mathcal{H}}_{G}\mathbf{D}_{\boldsymbol{\alpha}}^{-1}$ is a positive definite matrix. In analogous way, we prove that for any bounded continuous function $\psi(\mathbf{x}):\mathbb{R}^{n}\rightarrow\mathbb{R}$:

\begin{equation}\label{dadimostrare2}
	\dfrac{\displaystyle{\int_{\mathbb{R}^{n}}} \exp\bigg(\!-NG\Big(\dfrac{\mathbf{x}}{\sqrt{\mathbf{N}}}+\boldsymbol{\mu}\Big)\bigg)\psi(\mathbf{x})d\mathbf{x}}{\displaystyle{\int_{\mathbb{R}^{n}}}\exp\bigg(\!-NG\Big(\dfrac{\mathbf{x}}{\sqrt{\mathbf{N}}}+\boldsymbol{\mu}\Big)\bigg)d\mathbf{x}}
	\rightarrow \bigg(\dfrac{\det \boldsymbol{\widetilde{\mathcal{H}}}_{G}(\boldsymbol{\mu})}{(2\pi)^{n}}\bigg)^{\frac{1}{2}}\int_{\mathbb{R}^{n}} \exp\Big(-\frac{1}{2}\langle\boldsymbol{\widetilde{\mathcal{H}}}_{G}(\boldsymbol{\mu})\mathbf{x},\mathbf{x}\rangle\Big)\psi(\mathbf{x})d\mathbf{x}.
	\end{equation}
	\noindent  The multivariate Gaussian obtained is the convolution of the distribution of the random vector $(W_{1},\dots,W_{n})$ with the distribution of the random vector $\bar{\mathbf{S}}^{1}(\boldsymbol{\sigma})$ given by (\ref{vettore.S.1}). Indicated with $\phi_{\mathbf{W}}(\boldsymbol{\lambda})$, $\phi_{\bar{\mathbf{S}}^{1}}(\boldsymbol{\lambda})$ and $\phi_{\mathbf{W}+\bar{\mathbf{S}}^{1}}(\boldsymbol{\lambda})$ respectively the characteristic function of the random vectors $(W_{1},\dots,W_{n})$, of the random vector (\ref{vettore.S.1}) and of their sum the following equality holds:
	\begin{equation}\label{prodcaratteristiche}
	\phi_{\mathbf{W}+\bar{\mathbf{S}}^{1}}(\boldsymbol{\lambda})=\phi_{\mathbf{W}}(\boldsymbol{\lambda})\;\phi_{\bar{\mathbf{S}}^{1}}(\boldsymbol{\lambda})\;.
	\end{equation}
	We remember that the characteristic function of a random vector $(X_{1},\dots,X_{n})$ whose joint distribution is a multivariate Gaussian $$\sqrt{\frac{\det \mathbf{B}}{(2\pi)^{n}}}\; \exp\Big(\!-\frac{1}{2}\langle\mathbf{B}\mathbf{x},\mathbf{x}\rangle\Big)$$ is $\phi(\boldsymbol{\lambda})=\exp(-1/2\langle\mathbf{B}^{-1}\boldsymbol{\lambda},\boldsymbol{\lambda}\rangle)$ where $\mathbf{B}^{-1}$ is the covariance matrix of the vector $(X_{1},\dots,X_{n})$. Thus the equality (\ref{prodcaratteristiche}) allows to determine the covariance matrix of the vector $\bar{\mathbf{S}}^{1}(\boldsymbol{\sigma})$ taking off the matrix $\mathbf{A}^{-1}$ from $\boldsymbol{\widetilde{\mathcal{H}}}_{G}^{-1}$. By calculus it is easy to verify that 
\begin{equation}	
\boldsymbol{\widetilde{\mathcal{H}}}_{G}^{-1}-\mathbf{A}^{-1}=\boldsymbol{\widetilde{\chi}}. 
\end{equation}

To complete the proof we have to show that the matrix $\boldsymbol{\widetilde{\chi}}$ is positive definite. Consider the strictly convex function
\begin{equation}\label{funzione.fi.multi}
\Phi(x_{1},\dots,x_{n})=\sum_{l=1}^{n}\alpha_{l}\ln\bigg(\!\cosh\bigg(\sum_{s=1}^{n}\alpha_{s}J_{ls}x_{s}+h_{l}\bigg)\bigg)=\frac{1}{2}\sum_{l, s=1}^{n}\alpha_{l}\alpha_{s}J_{ls}x_{l}x_{s}-G(x_{1},\dots,x_{n})
\end{equation}
we can write $\boldsymbol{\mathcal{H}}_{G}=\mathbf{D}_{\boldsymbol{\alpha}}\mathbf{D}_{\boldsymbol{\alpha}}\mathbf{J}\mathbf{D}_{\boldsymbol{\alpha}}\mathbf{D}_{\boldsymbol{\alpha}}-\boldsymbol{\mathcal{H}}_{\Phi}$ where $\boldsymbol{\mathcal{H}}_{\Phi}$ is the Hessian matrix of the function $\Phi$.
Since $\mathbf{A}=\mathbf{D}_{\boldsymbol{\alpha}}\mathbf{J}\mathbf{D}_{\boldsymbol{\alpha}}$ we get  $\boldsymbol{\widetilde{\mathcal{H}}}_{G}=\mathbf{A}-\boldsymbol{\widetilde{\mathcal{H}}}_{\Phi}$ where $\boldsymbol{\widetilde{\mathcal{H}}}_{\Phi}=\mathbf{D}_{\boldsymbol{\alpha}}^{-1}\boldsymbol{\mathcal{H}}_{\Phi}\mathbf{D}_{\boldsymbol{\alpha}}^{-1}$.
Multiplying $\boldsymbol{\widetilde{\chi}}$ by the positive definite matrix $\boldsymbol{\widetilde{\mathcal{H}}}_{G}$ we obtain
\begin{equation}\label{convessita.fi.multi}
 \boldsymbol{\widetilde{\mathcal{H}}}_{G}(\boldsymbol{\mu})\boldsymbol{\widetilde{\chi}}=\Big(\mathbf{A}-\boldsymbol{\widetilde{\mathcal{H}}}_{\Phi}(\boldsymbol{\mu})\Big)\Big(\Big(\mathbf{A}-\boldsymbol{\widetilde{\mathcal{H}}}_{\Phi}(\boldsymbol{\mu})\Big)^{-1}-\mathbf{A}^{-1}\Big)=\boldsymbol{\widetilde{\mathcal{H}}}_{\Phi}(\boldsymbol{\mu})\mathbf{A}^{-1}.
\end{equation}
Since all matrices involved in (\ref{convessita.fi.multi}) are symmetric and $\boldsymbol{\widetilde{\mathcal{H}}}_{\Phi}(\boldsymbol{\mu})\mathbf{A}^{-1}$ is positive define it follows that also $\boldsymbol{\widetilde{\chi}}$ is positive definite.
Hence the random vector (\ref{vettore.S.1}), converges to a multivariate Gaussian which covariance matrix is $\boldsymbol{\widetilde{\chi}}$.

\subsection{Proof of Theorem \ref{teo2}}
In order to prove theorem \ref{teo2} we introduce the conditional joint distribution (see \cite{ellis1990limit}) of a configuration $\boldsymbol{\sigma}$ conditioned on the event $(m_{1}(\boldsymbol{\sigma}),\dots,m_{n}(\boldsymbol{\sigma}))\in B(\boldsymbol{\mu},d)$
\begin{equation}\label{BG.condizionata}
 P_{N,\mathbf{J},\mathbf{h},d}\{\boldsymbol{\sigma}\}=\frac{1}{Z_{N}(\mathbf{J},\mathbf{h},d)}\exp\Big(\frac{1}{2N}\langle\mathbf{J}\mathbf{s},\mathbf{s}\rangle+\langle\mathbf{h},\mathbf{s}\rangle\Big)1_{B(\boldsymbol{\mu},d)}\Big(\frac{\mathbf{s}}{\mathbf{N}}\Big)d\nu_{S}(\mathbf{s})
\end{equation}
where $1_{B(\boldsymbol{\mu},d)}$ is the indicator function of the ball $B(\boldsymbol{\mu},d)$, 
$\nu_{S}$ denotes the distribution of the random vector $(S_{1}(\boldsymbol{\sigma}),\dots,S_{n}(\boldsymbol{\sigma}))$ on $(\mathbb{R}^{N},\prod_{i=1}^{N}\rho(\sigma_{i}))$ and $Z_{N}(\mathbf{J},\mathbf{h},d)$ is the normalizing constant.\\
To prove the theorem we need the following lemmas.
\begin{lemma}\label{lemma.multi.4}
Let $\boldsymbol{\mu}$ be a global minimum point of the function $G$ given by (\ref{funzioneG}). Let $k$ be the homogeneous type of $\boldsymbol{\mu}$. Define $f=\min\{G(\mathbf{x})|\mathbf{x}\in\mathbb{R}^{n}\}$. Then there exists a positive number $\delta_{\boldsymbol{\mu}}$ such that for any $\mathbf{t}\in\mathbb{R}^{n}$, any $k\in\mathbb{N}$, any $\delta\in(0,\delta_{\boldsymbol{\mu}}]$ and any bounded continuous function $\phi:\mathbb{R}^{n}\rightarrow\mathbb{R}$
\begin{multline}\label{tesi.lemma.multi.4}
\lim_{N\rightarrow\infty}e^{-\langle\mathbf{t},\mathbf{N}^{1/2k}\boldsymbol{\mu}\rangle}\Big(\prod_{l=1}^{n}N_{l}\Big)^{1/2k}e^{Nf}\int_{B(\boldsymbol{\mu},\delta)}e^{-NG(\mathbf{x})+\langle\mathbf{t},\mathbf{N}^{1/2k}\mathbf{x}}\rangle\phi(\mathbf{x})d\mathbf{x}\\=
\phi(\boldsymbol{\mu})\int_{\mathbb{R}^{n}}\exp\bigg(-G^{\boldsymbol{\mu}}_{2k}\Big(\dfrac{\mathbf{x}}{\boldsymbol{\alpha}^{1/2k}}\Big)+\langle\mathbf{t},\mathbf{x}\rangle\bigg)d\mathbf{x} 
\end{multline}
\end{lemma}
See appendix for the proof.

\begin{lemma}\label{lemma.multi.5}
Let $P_{N,\mathbf{J},\mathbf{h},d}\{\boldsymbol{\sigma}\}$ be the joint distribution of $\boldsymbol{\sigma}=(\sigma_{1},\dots,\sigma_{N})$. Let $V_{\gamma}$ be the random vector
\begin{equation}\label{vettorone}
	\bigg(\dfrac{W_{1}}{(N_{1})^{1/2-\gamma}},\dots,\dfrac{W_{n}}{(N_{n})^{1/2-\gamma}}\bigg)+\bigg(\dfrac{S_{1}(\boldsymbol{\sigma})-N_{1}\mu_{1}}{(N_{1})^{1-\gamma}},\dots, \dfrac{S_{n}(\boldsymbol{\sigma})-N_{n}\mu_{n}}{(N_{n})^{1-\gamma}}\bigg)
	\end{equation}
where $\mathbf{W}\sim N(0,\mathbf{A}^{-1})$ and $\mathbf{A}=\mathbf{D}_{\boldsymbol{\alpha}}\mathbf{J}\mathbf{D}_{\boldsymbol{\alpha}}$ is a positive definite matrix.\\ 
Then
\begin{equation}\label{tesi.lemma.multi.5}
\Big\langle e^{\langle\mathbf{t},\mathbf{V}_{\gamma}\rangle}\Big\rangle_{BG_{d}}=
\dfrac{e^{-\langle\mathbf{t},\mathbf{N}^{\gamma}\boldsymbol{\mu}\rangle}\displaystyle{\int_{\mathbb{R}^{n}}}\exp\bigg(-\frac{N}{2}\langle\widetilde{\mathbf{J}}\mathbf{x},\mathbf{x}\rangle+\langle\mathbf{t},\mathbf{N}^{\gamma}\mathbf{x}\rangle\bigg)I_{N}(\mathbf{x},\boldsymbol{\mu},d)d\mathbf{x}}{\displaystyle{\int_{\mathbb{R}^{n}}}\exp\bigg(-\frac{N}{2}\langle\widetilde{\mathbf{J}}\mathbf{x},\mathbf{x}\rangle+\langle\mathbf{t},\mathbf{N}^{\gamma}\mathbf{x}\rangle\bigg)I_{N}(\mathbf{x},\boldsymbol{\mu},d)d\mathbf{x}}\end{equation}
where $\langle\cdot\rangle_{BG_{d}}$ denotes the expectation value with respect to the conditional distribution (\ref{BG.condizionata}) and
\begin{equation}
 I_{N}(\mathbf{x},\boldsymbol{\mu},d)=\int_{\{\frac{\mathbf{s}}{\mathbf{N}}\in B(\boldsymbol{\mu},d)\}}\exp(\langle\mathbf{J}\boldsymbol{\alpha}\mathbf{x},\mathbf{s}\rangle+\langle\mathbf{h},\mathbf{s}\rangle)d\nu_{S}(\mathbf{s}).
\end{equation}
\end{lemma}
See appendix for the proof.

We give the proof of theorem \ref{teo2} for $\boldsymbol{\mu}=\boldsymbol{\mu}^{1}$. The other global minimum points are handled identically. Fix $\mathbf{t}\in\mathbb{R}^{n}$ we choose the number $\delta_{p}>0$, $p=1,\dots,P$ according to lemma \ref{lemma.multi.4}. For all $\delta\in(0,\delta_{p}]$
\begin{multline}\label{iniziamente}
\lim_{N\rightarrow\infty}e^{-\langle\mathbf{t},\mathbf{N}^{1/2k_{p}}\boldsymbol{\mu}^{p}\rangle}\Big(\prod_{l=1}^{n}N_{l}\Big)^{1/2k_{p}}e^{Nf}\int_{B(\boldsymbol{\mu}^{p},\delta)}e^{-NG(\mathbf{x})+\langle\mathbf{t},\mathbf{N}^{1/2k_{p}}\mathbf{x}\rangle}d\mathbf{x}\\
=\int_{\mathbb{R}^{n}}\exp\bigg(-G^{\boldsymbol{\mu}^{p}}_{2k_{p}}\Big(\frac{\mathbf{x}}{\boldsymbol{\alpha}^{1/2k_{p}}}\Big)+\langle\mathbf{t},\mathbf{x}\rangle\bigg)d\mathbf{x}
\end{multline}
Define
\begin{equation}
I_{N}^{c}(\mathbf{x},\boldsymbol{\mu},d)=\int_{\{\frac{\mathbf{s}}{\mathbf{N}}\in B^{c}(\boldsymbol{\mu},d)\}}\exp(\langle\mathbf{J}\boldsymbol{\alpha}\mathbf{x},\mathbf{s}\rangle+\langle\mathbf{h},\mathbf{s}\rangle)d\nu_{S}(\mathbf{s}).
\end{equation}
It is easy to verify that:
\begin{align}\label{tortainarrivo}
I_{N}(\mathbf{x},\boldsymbol{\mu},d)+I_{N}^{c}(\mathbf{x},\boldsymbol{\mu},d)&=\exp(N\Phi(\mathbf{x})).
\end{align}
where the function $\Phi$ is given by (\ref{funzione.fi.multi}).
For any $p=1,\dots,P$, any $0<\theta<\bar{\delta}$ and any $0<\delta\leq\delta_{p}$, define
\begin{multline}
K_{N}(\mathbf{t},\delta,\boldsymbol{\mu}^{p},\theta,k_{p})=\Big(\prod_{l=1}^{n}N_{l}\Big)^{1/2k_{p}}\exp(Nf-\langle\mathbf{t},\mathbf{N}^{1/2k_{p}}\boldsymbol{\mu}^{p}\rangle)\\
\times\int_{B(\boldsymbol{\mu},\delta)}\!\!\!\exp\Big(-\frac{N}{2}\langle\widetilde{\mathbf{J}}\mathbf{x},\mathbf{x}\rangle+\langle\mathbf{t},\mathbf{N}^{1/2k_{p}}\mathbf{x}\rangle\Big)I_{N}^{c}(\mathbf{x},\boldsymbol{\mu}^{p},\theta)d\mathbf{x}
\end{multline}
Since $\mathbb{R}^{n}=B(\boldsymbol{\mu}^{1},\delta_{1})\cup B^{c}(\boldsymbol{\mu}^{1},\delta_{1})$ and $I_{N}(\mathbf{x},\boldsymbol{\mu}^{1},d)=e^{N\Phi(\mathbf{x})}-I_{N}^{c}(\mathbf{x},\boldsymbol{\mu}^{1},d)$ by lemma \ref{lemma.multi.5} (with $\gamma=1/2k_{1}$) after have
multiplied numerator and denominator by $(\prod_{l=1}^{n}N_{l})^{1/2k_{1}}e^{Nf}$ we obtain
\begin{equation}
\langle e^{\langle\mathbf{t},\mathbf{V}_{1/2k_{1}}\rangle}\rangle_{BG_{d}}=
\dfrac{L_{N}(\mathbf{t},\delta_{1},\boldsymbol{\mu}^{1},k_{1})-K_{N}(\mathbf{t},\delta_{1},\boldsymbol{\mu}^{1},d,k_{1})+M_{N}(\mathbf{t},\delta_{1},\boldsymbol{\mu}^{1},d,k_{1})}{L_{N}(\mathbf{0},\delta_{1},\boldsymbol{\mu}^{1},k_{1})-K_{N}(\mathbf{0},\delta_{1},\boldsymbol{\mu}^{1},d,k_{1})+M_{N}(\mathbf{0},\delta_{1},\boldsymbol{\mu}^{1},d,k_{1})}
\end{equation}
where the random vector $\mathbf{V}_{1/2k_{1}}$ is defined by (\ref{vettorone}), 
\begin{align}
L_{N}(\mathbf{t},\delta_{1},\boldsymbol{\mu}^{1},k_{1})&=\Big(\prod_{l=1}^{n}N_{l}\Big)^{1/2k_{1}}\exp(Nf-\langle\mathbf{t},\mathbf{N}^{1/2k_{1}}\boldsymbol{\mu}^{1}\rangle)\nonumber\\
&\quad\times\int_{B(\boldsymbol{\mu}^{1},\delta_{1})}\exp\Big(-\frac{N}{2}\langle\widetilde{\mathbf{J}}\mathbf{x},\mathbf{x}\rangle+\langle\mathbf{t},\mathbf{N}^{1/2k_{1}}\mathbf{x}\rangle+N\Phi(\mathbf{x})\Big)d\mathbf{x}\nonumber\\
&=\Big(\prod_{l=1}^{n}N_{l}\Big)^{1/2k_{1}}\exp(Nf-\langle\mathbf{t},\mathbf{N}^{1/2k_{1}}\boldsymbol{\mu}^{1}\rangle)\nonumber\\
&\quad\times\int_{B^{c}(\boldsymbol{\mu}^{1},\delta_{1})}\exp(-NG(\mathbf{x})+\langle\mathbf{t},\mathbf{N}^{1/2k_{1}}\mathbf{x}\rangle)d\mathbf{x}
\end{align}
and 
\begin{align}
M_{N}(\mathbf{t},\delta_{1},\boldsymbol{\mu}^{1},d,k_{1})&=\Big(\prod_{l=1}^{n}N_{l}\Big)^{1/2k_{1}}\exp(Nf-\langle\mathbf{t},\mathbf{N}^{1/2k_{1}}\boldsymbol{\mu}^{1}\rangle)\nonumber\\
&\quad\times\int_{B(\boldsymbol{\mu}^{1},\delta_{1})}\!\!\!\!\!\exp\Big(-\frac{N}{2}\langle\widetilde{\mathbf{J}}\mathbf{x},\mathbf{x}\rangle+\langle\mathbf{t},\mathbf{N}^{1/2k_{1}}\mathbf{x}\rangle\Big)I_{N}(\mathbf{x},\boldsymbol{\mu}^{1},d)d\mathbf{x}.
\end{align}
To proceed we need the following:
\begin{lemma}\label{ultimo.lemma}
Let $\boldsymbol{\mu}^{1},\dots,\boldsymbol{\mu}^{P}$ be minima points of the function $G$ given by (\ref{funzioneG}). Let $k_{1},\dots,k_{p}$ be their homogeneous type.
For any $\theta>0$ there exists $\epsilon>0$ such that for each $p=1,\dots,P$
\begin{equation}\label{tesi.lemma.funzione.k}
 K_{N}(\mathbf{t},\delta_{p},\boldsymbol{\mu}^{p},\theta,k_{p})=O(e^{-N\epsilon}) \quad\text{as}\;N\rightarrow\infty
\end{equation}
\end{lemma}
The proof of the lemma is in the appendix.
By (\ref{ultimo.lemma}) with $p=1$ and $\theta=d$ there exists $\epsilon_{1}>0$ such that
\begin{equation}\label{cavolini}
 K_{N}(\mathbf{t},\delta_{1},\boldsymbol{\mu}^{1},d,k_{1})=O(e^{-N\epsilon_{1}}).
\end{equation}
Now we will prove that there exists also $\epsilon_{2}>0$ such that
\begin{equation}\label{checavoli}
 M_{N}(\mathbf{t},\delta_{1},\boldsymbol{\mu}^{1},d,k_{1})=O(e^{-N\epsilon_{2}}).
\end{equation}
Define the closet set 
\begin{equation}
 V=\mathbb{R}^{n}-\bigcup_{p=1}^{P}B(\boldsymbol{\mu}^{p},\delta_{p})
\end{equation}
Then 
\begin{equation}
 B^{c}(\boldsymbol{\mu}^{1},\delta_{1})\subset V\cup\bigcup_{p=2}^{P}B(\boldsymbol{\mu}^{p},\delta_{p}).
\end{equation}
Thus we can write
\begin{align}
M_{N}(\mathbf{t},\delta_{1},&\boldsymbol{\mu}^{1},d,k_{1})\nonumber\\
&=\Big(\prod_{l=1}^{n}N_{l}\Big)^{1/2k_{1}}\exp(Nf-\langle\mathbf{t},\mathbf{N}^{1/2k_{1}}\boldsymbol{\mu}^{1}\rangle)\nonumber\\
&\quad\times\int_{V\cup\bigcup_{p=2}^{P}B(\boldsymbol{\mu}^{p},\delta_{p})}\!\!\!\!\!\exp\Big(-\frac{N}{2}\langle\widetilde{\mathbf{J}}\mathbf{x},\mathbf{x}\rangle+\langle\mathbf{t},\mathbf{N}^{1/2k_{1}}\mathbf{x}\rangle\Big)I_{N}(\mathbf{x},\boldsymbol{\mu}^{1},d)d\mathbf{x}
\end{align}
Since $d<\bar{\delta}$ and $||\boldsymbol{\mu}^{1}-\boldsymbol{\mu}^{p}||\geq\bar{\delta}$ for $p=2,\dots,P$ we have $B(\boldsymbol{\mu}^{1},\delta_{1})\subset B^{c}(\boldsymbol{\mu}^{p},\bar{\delta}-d)$ hence for each $\mathbf{x}\in\mathbb{R}^{n}$ and $p=2,\dots,P$
\begin{equation}\label{quasiletto}
 I_{N}(\mathbf{x},\boldsymbol{\mu}^{1},d)\leq I_{N}^{c}(\mathbf{x},\boldsymbol{\mu}^{p},\bar{\delta}-d).
\end{equation}
Moreover by (\ref{tortainarrivo})
\begin{equation}\label{letto}
I_{N}(\mathbf{x},\boldsymbol{\mu}^{1},d)\leq\exp(N\Phi(\mathbf{x}))
\end{equation}
Using (\ref{quasiletto}) in the integrals over $B(\boldsymbol{\mu}^{p},\delta_{p})$, $p=2,\dots,P$ and (\ref{letto}) in the integral over $V$ we obtain
\begin{align}\label{ultimissime cose}
M_{N}(\mathbf{t},\delta_{1},\boldsymbol{\mu}^{1},d,k_{1})&=\Big(\prod_{l=1}^{n}N_{l}\Big)^{1/2k_{1}}\exp(Nf-\langle\mathbf{t},\mathbf{N}^{1/2k_{1}}\boldsymbol{\mu}^{1}\rangle)\nonumber\\
&\quad\times\int_{V}\exp(-NG(\mathbf{x})+\langle\mathbf{t},\mathbf{N}^{1/2k_{1}}\mathbf{x}\rangle)d\mathbf{x}\nonumber\\
&\quad+\sum_{p=2}^{P}\exp\Big(\langle\mathbf{t},\mathbf{N}^{1/2k_{1}}(\boldsymbol{\mu}^{p}-\boldsymbol{\mu}^{1})\rangle\Big)K_{N}(\mathbf{t},\delta_{p},\boldsymbol{\mu}^{p},\bar{\delta}-d,k_{1})
\end{align}
Since the set $V$ does not contain minima points, for any $\mathbf{t}\in\mathbb{R}^{n}$ there exists $\epsilon>0$ such that
\begin{equation}\label{tesi.lemma.3.multi}
e^{Nf}\int_{V} \exp(-NG(\mathbf{x})+\langle \mathbf{t},\mathbf{N}^{1/2k_{1}}\mathbf{x}\rangle)d\mathbf{x}=O(e^{-N\epsilon}) \quad\quad\quad N\rightarrow\infty.
\end{equation}
Applying (\ref{tesi.lemma.3.multi}) to the first term of the right-hand side of (\ref{ultimissime cose}) and lemma \ref{ultimo.lemma} to each term of the sum, the result (\ref{checavoli}) holds.

By (\ref{cavolini}) and (\ref{checavoli}) we have
\begin{equation}
 \lim_{N\rightarrow\infty}\langle e^{\langle\mathbf{t},\mathbf{V}_{1/2k_{1}}\rangle}\rangle_{BG_{d}}=\dfrac{L_{N}(\mathbf{t},\delta_{1},\boldsymbol{\mu}^{1},k_{1})}{L_{N}(\mathbf{0},\delta_{1},\boldsymbol{\mu}^{1},k_{1})}
=\dfrac{\displaystyle{\int_{\mathbb{R}^{n}}}\exp\bigg(-G_{2k_{1}}^{\boldsymbol{\mu}^{1}}\Big(\frac{\mathbf{x}}{\boldsymbol{\alpha}^{1/2k_{p}}}\Big)+\langle\mathbf{t},\mathbf{x}\rangle\bigg)d\mathbf{x}}{\displaystyle{\int_{\mathbb{R}^{n}}}\exp\bigg(-G_{2k_{p}}^{\boldsymbol{\mu}^{1}}\Big(\frac{\mathbf{x}}{\boldsymbol{\alpha}^{1/2k_{p}}}\Big)\bigg)d\mathbf{x}}
\end{equation}
where in the last identity we use (\ref{iniziamente}). By the assumption on the random vector $W$ the theorem \ref{teo2} is proved.

\section{Examples}
\noindent We now analyze the case of two populations of the same cardinality. The Hamiltonian 
\begin{equation}
H_{N}(m_{1},m_{2})=-\frac{N}{8}\Big(J_{11}m_{1}^{2}+J_{22}m_{2}^{2}+2J_{12}m_{1}m_{2}+4h_{1}m_{1}+4h_{2}m_{2}\Big)
\end{equation}
is a convex function of the magnetizations if the reduced interaction matrix $\mathbf{J}$ 
is positive definite, that is $J_{11}>0$ and $J_{11}J_{22}-J_{12}^{2}>0$. A stationary point $(\mu_{1},\mu_{2})$ of the function $G$
\begin{equation}
\begin{split}
G(x_{1},x_{2}) &=\frac{1}{8}\Big(J_{11}x_{1}^{2}+2J_{12}x_{1}x_{2}+J_{22}x_{2}^{2}\Big)
-\frac{1}{2}\ln\Big(\cosh\Big(\frac{J_{11}}{2}x_{1}+\frac{J_{12}}{2}x_{2}+h_{1}\Big)\Big)\\
&\quad-\frac{1}{2}\ln\Big(\cosh\Big(\frac{J_{12}}{2}x_{1}+\frac{J_{22}}{2}x_{2}+h_{2}\Big)\Big)
\end{split}
\end{equation}
is a minimum point of homogeneous type $k=1$ if:
\begin{equation}\label{condizioniminimo}
\begin{cases}
(\boldsymbol{\mathcal{H}}_{G})_{11}(\mu_{1},\mu_{2})>0 \\
\det_{\boldsymbol{\mathcal{H}}_{G}}(\mu_{1},\mu_{2})>0
\end{cases} 
\end{equation}
where:
\begin{equation}
(\boldsymbol{\mathcal{H}}_{G})_{11}(\mu_{1},\mu_{2})=\frac{1}{8}\Big(2J_{11}-J_{11}^{2}(1-\mu_{1}^{2})-J_{12}^{2}(1-\mu_{2}^{2})\Big)
\end{equation}
\noindent and the determinant is:
\begin{equation}
\det_{\boldsymbol{\mathcal{H}}_{G}}(\mu_{1},\mu_{2})
=\frac{\det\mathbf{J}}{64}\Big(4-2J_{11}(1-\mu_{1}^{2})-2J_{22}(1-\mu_{2}^{2}) +\det\mathbf{J}(1-\mu_{1}^{2})(1-\mu_{2}^{2})\Big).
\end{equation}
For example if we consider the particular case in which the external field $h_{1}$ and $h_{2}$ are equal to zero and the parameters $J_{11}$ and $J_{22}$ are the same, the stationary point $(0,0)$ verifies the conditions (\ref{condizioniminimo}) if:
\begin{equation}
\begin{cases}
0<J_{11}\leq 1\\
 -J_{11}<J_{12}<J_{11}
 \end{cases}
\quad\cup\quad\quad
\begin{cases}
1<J_{11}<2\\
J_{11}-2<J_{12}<2-J_{11}\;.
\end{cases}
\end{equation}
Thus for these choices of the parameters the random vector
\begin{equation}
\bigg(\dfrac{S_{1}(\boldsymbol{\sigma})}{\sqrt{N_{1}}},\dfrac{S_{2}(\boldsymbol{\sigma})}{\sqrt{N_{2}}}\bigg)
\end{equation}
weakly converges to a bivariate normal distribution. To have a minimum point of homogeneous type $k>1$ the Hessian matrix of $G$ computed in minimum point $(\mu_{1},\mu_{2})$ must be equal to the matrix
with zero elements. This condition means:
\begin{equation}
\begin{cases}
J_{11}\geq2\\
J_{22}\geq2\\
J_{12}=0\\
\mu_{1}^{2}=\dfrac{J_{11}-2}{J_{11}}\\
\mu_{2}^{2}=\dfrac{J_{22}-2}{J_{22}}\;.
\end{cases} 
\end{equation}
Only if the third partial derivatives of $G$ computed in $(\mu_{1},\mu_{2})$ are equal to zero the point can be a minimum point. This is verified if and only if $J_{11}=J_{22}=2$. Hence $(\mu_{1},\mu_{2})=(0,0)$. Computing the partial derivatives of fourth order we can assert that $(0,0)$ is a minimum point  of homogeneous type $k=2$. Thus the limiting distribution of the random vector
\begin{equation}
\bigg(\dfrac{S_{1}(\boldsymbol{\sigma})}{N_{1}^{3/4}},\dfrac{S_{2}(\boldsymbol{\sigma})}{N_{2}^{3/4}}\bigg)
\end{equation}
is proportional to $e^{-1/24(x_{1}^{4}+x_{4}^{2})}$.

As the parameter $J_{12}=0$, we are also able to describe the limiting distribution of the random vector $(S_{1}(\boldsymbol{\sigma}),S_{2}(\boldsymbol{\sigma}))$, properly normalized, beyond the homogeneity hypothesis on the minimum points of the function $G$. That is why in this case $S_{1}(\boldsymbol{\sigma})$ and $S_{2}(\boldsymbol{\sigma})$ are independent random variables and thus their joint distribution is the product of the marginal distributions.  We observe that as $J_{11}=2$ and $h_{1}=0$, for each value of the parameters $J_{22}$ and $h_{2}$, the determinant of the Hessian matrix of $G$ computed in the minimum point $(0,\mu_{2})$ is equal to zero. The type of the minimum point $(0,\mu_{2})$ is not homogeneous whenever the second partial derivative of $G$ with respect to $x_{2}$ is different from zero. This condition is verifies as $J_{22}\neq 2$ or $h_{2}\neq 0$. For these choices of the parameters, the limiting distribution of the random vector 
\begin{equation}
\bigg(\dfrac{S_{1}(\boldsymbol{\sigma})}{N_{1}^{3/4}},\dfrac{S_{2}(\boldsymbol{\sigma})}{\sqrt{N_{2}}}\bigg)
\end{equation}
is given by the product of a Gaussian distribution with an exponential distribution of the forth order.

\section{Conclusions and Outlooks}
In this paper we have generalized to multi-species Curie-Weiss models the study of the normalized sums
of spins and their limiting distributions. We worked under a condition of convexity of the reduced interaction 
matrix which allows us to use the Ellis-Newman method. The theorems presented in this work obtain a complete
classification of the distribution when the first non vanishing partial derivatives are all the same order
(homogeneity hypothesis). The extension to non convex interactions or the complete classification of the
limiting distribution beyond the homogeneity hypothesis will be subject of further investigation.


\appendix
\section{Proof of lemma \ref{proposizione}}
Considering the function 
\begin{equation}
\widetilde{G}(x_{1},\dots,x_{n})=\frac{1}{2}\sum_{l, s=1}^{n}\alpha_{l}\alpha_{s}J_{ls}x_{l}x_{s}-\sum_{l=1}^{n}\alpha_{l}t_{l}\Big(\sum_{s=1}^{n}\alpha_{s}J_{ls}\;x_{s}+h_{l}\Big)
\end{equation}
where $\mathbf{x}=(x_{1},\dots,x_{n})\in\mathbb{R}^{n}$ and $\mathbf{t}=(t_{1},\dots,t_{n})\in\{-1,+1\}^{n}$, the following inequality holds:
\begin{equation}\label{convessitaG}
G(\mathbf{x},\mathbf{t}) \geq \min_{\mathbf{t}}\widetilde{G}(\mathbf{x},\mathbf{t}).
\end{equation}
Since the reduced interaction matrix $\mathbf{J}$ is positive definite the function on the right-hand side of (\ref{convessitaG}) is a quadratic function with a positive definite Hessian matrix. Thus by inequality (\ref{convessitaG}) the global minimum points (at least one) of the function $G$ belong to a same compact level set. Moreover these points, solutions of the mean field equations (\ref{campomedio}), are isolated and the function $G$ is analytic. Then the number of global minimum points is finite.\\
By (\ref{convessitaG}) we have:
\begin{equation}\label{proffite}
\int_{\mathbb{R}^{n}}\exp(-NG(\mathbf{x}))d\mathbf{x}\leq\int_{\mathbb{R}^{n}}\exp\Big(-N\min_{\mathbf{t}}\widetilde{G}(\mathbf{x},\mathbf{t})\Big)d\mathbf{x}.
\end{equation}
\noindent Since the argument of the integral on the right hand side of inequality (\ref{proffite}) is a Gaussian density function the statement (\ref{proprietaG}) follows.

\section{Proof of lemma \ref{lemma2}}
Given $\theta_{1},\dots, \theta_{n}$ real\\
\begin{multline*}
\!\!P\bigg\{\!\dfrac{W_{1}}{(N_{1})^{1/2-\gamma}}+\dfrac{S_{1}(\boldsymbol{\sigma})\!-\!N_{1}m_{1}}{(N_{1})^{1-\gamma}}\!\leq\theta_{1},\!\dots ,\!\dfrac{W_{n}}{(N_{n})^{1/2-\gamma}}+\dfrac{S_{n}(\boldsymbol{\sigma})\!-\!N_{n}m_{n}}{(N_{n})^{1-\gamma}}\leq\theta_{n}\!\bigg\}\\\\\quad\quad=
P\;\Big\{\sqrt{N_{1}}W_{1}+S_{1}(\boldsymbol{\sigma})\in E_{1},\;\dots,\;\sqrt{N_{n}}W_{n}+S_{n}(\boldsymbol{\sigma})\in E_{n}\Big\}
\end{multline*}\\
where $E_{l}=(-\infty,\;(N_{l})^{1-\gamma}\theta_{l}+N_{l}m_{l}]$. The distribution of the random vector $(\sqrt{N_{1}}\;W_{1},\dots ,\sqrt{N_{n}}\;W_{n})$ is 
\begin{equation}\label{normale.bivariata.2}
\bigg(\frac{\det\widetilde{\mathbf{A}}}{(2\pi)^{n}}\bigg)^{\frac{1}{2}} \;\exp\Big(\!-\frac{1}{2}\langle\widetilde{\mathbf{A}}\mathbf{x},\mathbf{x}\rangle\Big)
\end{equation}
\noindent where it is easy to verify that $\widetilde{\mathbf{A}}=1/N\mathbf{J}$. We claim that since the matrix $\mathbf{A}$ is positive definite also $\widetilde{\mathbf{A}}$ has this property. 
The joint distribution of the random vector $(S_{1}(\boldsymbol{\sigma}),\dots ,S_{n}(\boldsymbol{\sigma}))$ is:\\
\begin{equation}\label{distribuzione.somme}
\frac{1}{Z_{N}(\mathbf{J},\mathbf{h})}\exp\bigg(\frac{1}{2N}\langle\mathbf{J}\mathbf{s},\mathbf{s}\rangle+\langle\mathbf{h},\mathbf{s}\rangle\bigg)d\nu_{S}(\mathbf{s})
\end{equation}\\
\noindent where $\nu_{S}(\mathbf{s})$ is the distribution of $(S_{1}(\boldsymbol{\sigma}),\dots ,S_{n}(\boldsymbol{\sigma}))$ on $(\mathbb{R}^{N},\prod_{i=1}^{N}\rho(\sigma_{i}))$. 
The distribution of the random vector (\ref{sommavettori}) is given by the convolution of the distribution (\ref{normale.bivariata.2}) with the distribution (\ref{distribuzione.somme}).
\noindent Thus:
\begin{align*}
P&\Big\{\sqrt{N_{1}}\;W_{1}+S_{1}(\boldsymbol{\sigma})\in E_{1},\dots ,\sqrt{N_{n}}\;W_{n}+S_{n}(\boldsymbol{\sigma})\in E_{n}\Big\} \nonumber\\\nonumber\\
&=\frac{1}{Z_{N}(\mathbf{J},\mathbf{h})}\bigg(\frac{\det\widetilde{\mathbf{A}}}{(2\pi)^{n}}\bigg)^{\frac{1}{2}}\nonumber\\
&\quad\times\!\!\iint_{\bigotimes\limits_{l=1}^{n} E_{l}\times\mathbb{R}^{n}}\!\!\!\!\!\!\exp\bigg(\!\frac{1}{2N}\Big(\!-\langle\mathbf{J}(\mathbf{w}-\mathbf{s}),(\mathbf{w}-\mathbf{s})\rangle+\langle\mathbf{J}\mathbf{s},\mathbf{s}\rangle\Big)+\langle\mathbf{h},\mathbf{s}\rangle\bigg)d\nu_{S}(\mathbf{s})d\mathbf{x}\nonumber\\\nonumber\\
&=\frac{1}{Z_{N}(\mathbf{J},\mathbf{h})}\bigg(\frac{\det\widetilde{\mathbf{A}}}{(2\pi)^{n}}\bigg)^{\frac{1}{2}}\nonumber\\
&\quad\times\!\!\int_{\bigotimes\limits_{l=1}^{n} E_{l}}\!\!\!\!\!\exp\bigg(\!\!-\frac{1}{2N}\langle\mathbf{J}\mathbf{w},\mathbf{w}\rangle\bigg)\!\!\int_{\mathbb{R}^{n}}\exp\bigg(\frac{1}{N}\langle\mathbf{J}\mathbf{w},\mathbf{s}\rangle+\langle\mathbf{h},\mathbf{s}\rangle\bigg)d\nu_{S}(\mathbf{s})d\mathbf{w}
\end{align*}
where $\bigotimes\limits_{l=1}^{n} E_{l}$ denotes the Cartesian product of the sets $E_{l}$. 

\noindent Since
\begin{align*}
\int_{\mathbb{R}^{n}}\exp\bigg(\frac{1}{N}\langle\mathbf{J}&\mathbf{w},\mathbf{s}\rangle+\langle\mathbf{h},\mathbf{s}\rangle\bigg)d\nu_{S}(\mathbf{s})\\&=\prod_{l=1}^{n}\int_{\mathbb{R}^{N_{1}}}\!\!\!\exp\bigg(\sum_{i\in P_{l}}\sigma_{i}\bigg(h_{l}+\frac{1}{N}\sum_{p=1}^{n}J_{lp}w_{p}\bigg)\bigg)\prod_{i\in P_{l}}d\rho(\sigma_{i})\nonumber\\
&=\prod_{l=1}^{n}\prod_{i\in P_{l}}\int_{\mathbb{R}}\exp\bigg(\sigma_{i}\bigg(h_{l}+\frac{1}{N}\sum_{p=1}^{n}J_{lp}w_{p}\bigg)\bigg)d\rho(\sigma_{i})
\end{align*}
making the following change of variables
\begin{equation*}
x_{l}=\dfrac{w_{l}-N_{l}m_{l}}{(N_{l})^{1-\gamma}}\quad\quad l=1,\dots, n
\end{equation*}
\noindent and integrating over $\mathbf{s}$, we obtain:

\begin{align}\label{calcoloneproseguo}
P&\Big\{\sqrt{N_{1}}\;W_{1}+S_{1}(\boldsymbol{\sigma})\in E_{1},\dots ,\sqrt{N_{n}}\;W_{n}+S_{n}(\boldsymbol{\sigma})\in E_{n}\Big\}\nonumber\\\nonumber\\
&=\dfrac{\prod\limits_{l=1}^{n}(N_{l})^{1-\gamma}}{Z_{N}(\mathbf{J},\mathbf{h})}\bigg(\frac{\det\widetilde{\mathbf{A}}}{(2\pi)^{n}}\bigg)^{\frac{1}{2}}\nonumber\\
&\quad\times\int_{-\infty}^{\theta_{1}}\dots\int\limits_{-\infty}^{\theta_{n}}\exp\bigg(\!-\frac{N}{2}\sum_{l, p=1}^{n}\alpha_{l}\alpha_{p}J_{lp}\Big(\dfrac{x_{l}}{N_{l}^{\;\gamma}}+m_{l}\Big)\Big(\dfrac{x_{p}}{N_{p}^{\;\gamma}}+m_{p}\Big)+\nonumber\\
&\quad+\sum_{l=1}^{n}N_{l}\ln\Big(\cosh\Big(h_{l}\sum_{p=1}^{n}\alpha_{p}J_{lp}\Big(\dfrac{x_{p}}{N_{p}^{\;\gamma}}+m_{p}\Big)\Big)\Big)\bigg)dx_{1}\dots dx_{n}\nonumber\\\nonumber\\
&=\dfrac{\prod\limits_{l=1}^{n}(N_{l})^{1-\gamma}}{Z_{N}(\mathbf{J},\mathbf{h})}\bigg(\frac{\det\widetilde{\mathbf{A}}}{(2\pi)^{n}}\bigg)^{\frac{1}{2}}\nonumber\\
&\quad\times\int_{-\infty}^{\theta_{1}}\dots\int\limits_{-\infty}^{\theta_{n}}\exp\bigg(\!-NG\Big(\dfrac{x_{1}}{N_{1}^{\;\gamma}}+m_{1},\dots,\dfrac{x_{n}}{N_{n}^{\;\gamma}}+m_{n}\Big)\bigg)dx_{1}\dots dx_{n}.
\end{align}\\
Taking $\theta_{1}\rightarrow\infty,\dots,\theta_{n}\rightarrow\infty$ the (\ref{calcoloneproseguo}) gives an equation for $Z_{N}(\mathbf{J},\mathbf{h})$ which when substituted back yields the result (\ref{distribuzione.tesi}). The integral in the last expression is finite by (\ref{proprietaG}).

\section{Proof of lemma \ref{lemma.multi.4}}
To easy the notation we define $\gamma=1/2k$. Making the change of variable
	\begin{equation}
	x_{l}=\mu_{l}+\dfrac{u_{l}}{N_{l}^{\gamma}}\quad l=1,\dots,n
	\end{equation}
	the left-hand side of (\ref{tesi.lemma.multi.4}) becomes
	\begin{equation}\label{passaggio.lemma.multi.4}
	\lim_{N\rightarrow\infty}\int_{||\frac{\mathbf{u}}{\mathbf{N}^{\gamma}}||\leq d}\phi\Big(\boldsymbol{\mu}+\frac{\mathbf{u}}{\mathbf{N}^{\gamma}}\Big)\exp\bigg(NB\Big(\boldsymbol{\mu}+\frac{\mathbf{u}}{\mathbf{N}^{\gamma}};\boldsymbol{\mu}\Big)+\langle\mathbf{t},\mathbf{x}\rangle\bigg)d\mathbf{x}
	\end{equation}
	where $B$ is the function defined in (\ref{funzione.B.multi}). By the conditions expressed in (\ref{proprieta.B.multi}) and dominate convergence theorem the limit (\ref{passaggio.lemma.multi.4}) is equal to
	\begin{equation}\label{passaggio2.lemma.multi.4}
	\phi(\boldsymbol{\mu})\int_{\mathbb{R}^{n}}\exp\bigg(-G^{\boldsymbol{\mu}}_{2k}\Big(\dfrac{\mathbf{u}}{\boldsymbol{\alpha}^{1/2k}}\Big)+\langle\mathbf{t},\mathbf{u}\rangle\bigg)d\mathbf{u}
	\end{equation}
	Since $G^{\boldsymbol{\mu}^{p}}_{2k}(\frac{\mathbf{x}}{\boldsymbol{\alpha}^{1/2k}})>0$ for every $\mathbf{x}$ different from zero, the integral in (\ref{passaggio2.lemma.multi.4}) is finite. 
	This completes the proof of the lemma. 

\section{Proof of lemma \ref{lemma.multi.5}}
By following the same proof of lemma \ref{lemma2} we have that the distribution of the random vector $V_{\gamma}$ is given by
\begin{equation}
 \dfrac{\exp\Big(-\frac{N}{2}\langle\widetilde{\mathbf{J}}(\boldsymbol{\mu}+\frac{\mathbf{x}}{\mathbf{N}^{\gamma}}),\boldsymbol{\mu}+\frac{\mathbf{x}}{\mathbf{N}^{\gamma}}\rangle\Big)I_{N}(\boldsymbol{\mu}+\frac{\mathbf{x}}{\mathbf{N}^{\gamma}},\boldsymbol{\mu},d)d\mathbf{x}}{\displaystyle{\int_{\mathbb{R}^{n}}}\exp\Big(-\frac{N}{2}\langle\widetilde{\mathbf{J}}(\boldsymbol{\mu}+\frac{\mathbf{x}}{\mathbf{N}^{\gamma}}),\boldsymbol{\mu}+\frac{\mathbf{x}}{\mathbf{N}^{\gamma}}\rangle\Big)d\mathbf{x}}
\end{equation}
where
\begin{equation}
I_{N}\Big(\boldsymbol{\mu}+\frac{\mathbf{x}}{\mathbf{N}^{\gamma}},\boldsymbol{\mu},d\Big)=\int_{\{\frac{\mathbf{s}}{\mathbf{N}}\in B(\boldsymbol{\mu},d)\}}\exp\Big(\langle \mathbf{J}\boldsymbol{\alpha}(\boldsymbol{\mu}+\frac{\mathbf{x}}{\mathbf{N}^{\gamma}}),\mathbf{s}\rangle+\langle \mathbf{h}, \mathbf{s}\rangle\Big)d\nu_{S}(\mathbf{s}).
\end{equation}
Thus 
\begin{equation}
\langle e^{\langle\mathbf{t},\mathbf{V}_{\gamma}\rangle}\rangle_{BG_{d}}=\dfrac{\displaystyle{\int_{\mathbb{R}^{n}}}e^{\langle\mathbf{t},\mathbf{x}\rangle}\exp\Big(-\frac{N}{2}\langle\widetilde{\mathbf{J}}(\boldsymbol{\mu}+\frac{\mathbf{x}}{\mathbf{N}^{\gamma}}),\boldsymbol{\mu}+\frac{\mathbf{x}}{\mathbf{N}^{\gamma}}\rangle\Big)I_{N}\Big(\boldsymbol{\mu}+\frac{\mathbf{x}}{\mathbf{N}^{\gamma}},\boldsymbol{\mu},d\Big)d\mathbf{x}}{\displaystyle{\int_{\mathbb{R}^{n}}}\exp\Big(-\frac{N}{2}\langle\widetilde{\mathbf{J}}(\boldsymbol{\mu}+\frac{\mathbf{x}}{\mathbf{N}^{\gamma}}),\boldsymbol{\mu}+\frac{\mathbf{x}}{\mathbf{N}^{\gamma}}\Big)d\mathbf{x}}
\end{equation}
Making the change of variable
\begin{equation}
u_{l}=\mu_{l}+\dfrac{x_{l}}{N_{l}^{\gamma}}\quad l=1,\dots,n
\end{equation}
the statement (\ref{tesi.lemma.multi.5}) holds.

\section{Proof of lemma \ref{ultimo.lemma}}
We prove the theorem for $p=1$. The proofs for other $p$ are similar. We observe that
\begin{align}
 \bigg\{\bigg(\frac{S_{1}(\boldsymbol{\sigma})}{N_{1}},\dots,&\frac{S_{n}(\boldsymbol{\sigma})}{N_{n}}\bigg)\in B^{c}(\boldsymbol{\mu}^{1},\theta)\bigg\}\nonumber\\
&\subset\bigcup_{l=1}^{n}\bigg\{\bigg|\frac{S_{l}(\boldsymbol{\sigma})}{N_{l}}-\mu_{l}^{1}\bigg|\geq \bar{\theta}\bigg\}\nonumber\\
&=\bigcup_{l=1}^{n}\bigg(\bigg\{\bigg|\frac{S_{l}(\boldsymbol{\sigma})}{N_{l}}\bigg|\leq\mu_{l}^{1}-\bar{\theta}\bigg\}\cup\bigg\{\bigg|\frac{S_{l}(\boldsymbol{\sigma})}{N_{l}}\bigg|\geq\mu_{l}^{1}+\bar{\theta}\bigg\}\bigg)
\end{align}
where $\bar{\theta}=\sqrt{n}\theta$. Thus
\begin{multline}\label{momento.te}
 I_{N}^{c}(\mathbf{x},\boldsymbol{\mu}^{1},\theta)\leq\sum_{l=1}^{n}\bigg(\int_{\{-\frac{s_{l}}{N_{l}}\geq-\mu_{l}^{1}+\bar{\theta}\}}\exp(\langle\mathbf{J}\boldsymbol{\alpha}\mathbf{x},\mathbf{s}\rangle+\langle\mathbf{h},\mathbf{s}\rangle)d\nu_{S}(\mathbf{s})\\
+\quad\int_{\{\frac{s_{l}}{N_{l}}\geq\mu_{l}^{1}+\bar{\theta}\}}\exp(\langle\mathbf{J}\boldsymbol{\alpha}\mathbf{x},\mathbf{s}\rangle+\langle\mathbf{h},\mathbf{s}\rangle)d\nu_{S}(\mathbf{s})\bigg)
\end{multline}
Consider one of the integrals of expression (\ref{momento.te})
\begin{align*}
\int_{\{\frac{s_{l}}{N_{l}}\geq\mu_{l}^{1}+\bar{\theta}\}}&\exp(\langle\mathbf{J}\boldsymbol{\alpha}\mathbf{x},\mathbf{s}\rangle+\langle\mathbf{h},\mathbf{s}\rangle)d\nu_{S}(\mathbf{s})\nonumber\\
&=\prod_{l=2}^{n}\cosh\Big(\sum_{q=1}^{n}\alpha_{q}J_{lq}x_{q}+h_{l}\Big)^{N_{l}}\nonumber\\
&\quad\times\int_{\{S_{1}(\boldsymbol{\sigma})\geq N_{1}(\mu_{1}^{1}+\bar{\theta})\}}\exp\Big(S_{1}(\boldsymbol{\sigma})\Big(\sum_{q=1}^{n}\alpha_{q}J_{1q}x_{q}+h_{1}\Big)\Big)\prod_{i\in P_{1}}d\rho(\sigma_{i}).
\end{align*}
By Chebishev's inequality for any $\tau>0$
\begin{align}\label{bamba1}
 \int_{\{S_{1}(\boldsymbol{\sigma})\geq N_{1}(\mu_{1}^{1}+\bar{\theta})\}}\!\!\!\!\exp&\Big(S_{1}(\boldsymbol{\sigma})\Big(\sum_{q=1}^{n}\alpha_{q}J_{1q}x_{q}+h_{1}\Big)\Big)\prod_{i\in P_{1}}d\rho(\sigma_{i})\nonumber\\
 &\leq\exp(-\alpha_{1}J_{11}\tau N_{1}(\mu_{1}^{1}+\bar{\theta}))\nonumber\\
 &\quad\times\int_{\mathbb{R}^{N_{1}}}\exp(\alpha_{1}J_{11}\tau\sum_{i\in P_{1}}\sigma_{i})\exp(\sum_{i\in P_{1}}\sigma_{i}(\alpha_{1}J_{11}x_{1}+\alpha_{2}J_{12}x_{2}+h_{1}))\nonumber\\
&=\exp(N_{1}(-\alpha_{1}J_{11}\tau(\mu_{1}^{1}+\bar{\theta})+\ln\cosh(\alpha_{1}J_{11}(x_{1}+\tau)+\alpha_{2}J_{12}x_{2}+h_{1}))).
\end{align}
By the mean field equations (\ref{campomedio}) we have:
\begin{equation}
 \frac{\partial}{\partial x_{1}}(\ln\cosh(\alpha_{1}J_{11}x_{1}+\alpha_{2}J_{12}x_{2}+h_{1}))(\boldsymbol{\mu}^{1})=\alpha_{1}J_{11}\mu_{1}^{1}.
\end{equation}
Thus we can choose $\delta>0$ and $\tau>0$ sufficiently small such that $\delta<\delta_{1}$ and 
\begin{equation}\label{bamba2}
\ln\cosh(\alpha_{1}J_{11}(x_{1}+\tau)+\alpha_{2}J_{12}x_{2}+h_{1})\\\leq\ln\cosh(\alpha_{1}J_{11}x_{1}+\alpha_{2}J_{12}x_{2}+h_{1})+\alpha_{1}J_{11}\mu_{1}^{1}\tau+\frac{1}{2}\alpha_{1}J_{11}\tau\bar{\theta}
\end{equation}
for each $\mathbf{x}\in B(\boldsymbol{\mu}^{1},\delta)$. The other integrals in (\ref{momento.te}) are handled in a similar way. 
At last applying the bounds (\ref{bamba1}) and (\ref{bamba2}) to (\ref{momento.te}), for all $\mathbf{x}\in B(\boldsymbol{\mu}^{1},\delta)$ we obtain:
\begin{equation}
I_{N}^{c}(\mathbf{x},\boldsymbol{\mu}^{1},\theta)\leq 2n\exp\Big(N\Big(-\frac{1}{2}\alpha_{1}^{2}J_{11}\tau\bar{\theta}+\Phi(\mathbf{x})\Big)\Big)
\end{equation}
where $\Phi$ is given by (\ref{funzione.fi.multi}). Hence
\begin{align}\label{pera}
K_{N}(\mathbf{t},\delta,\boldsymbol{\mu}^{1},\theta,k_{1})&\leq 2n (N_{l})^{1/2k_{1}}\exp\Big(Nf-\langle\mathbf{t},\mathbf{N}^{1/2k_{1}}\boldsymbol{\mu}^{1}\rangle-\frac{N}{2}\alpha_{1}^{2}J_{11}\tau\bar{\theta}\Big)\nonumber\\
&\quad\times\int_{B(\boldsymbol{\mu}^{1},\delta)}\exp(-NG(\mathbf{x})+\langle\mathbf{t},\mathbf{N}^{1/2k_{1}}\mathbf{x}\rangle)d\mathbf{x}.
\end{align}
Applying the lemma \ref{lemma.multi.4} to the expression on the right-hand side of (\ref{pera}) we obtain:
\begin{equation}\label{stima1}
K_{N}(\mathbf{t},\delta,\boldsymbol{\mu}^{1},\theta,k_{1})=O(e^{-\frac{N}{2}\alpha_{1}^{2}J_{11}\tau\bar{\theta}})\quad\text{as }N\rightarrow\infty.
\end{equation}
We now bound
\begin{align}
K_{N}(\mathbf{t},\delta_{1},&\boldsymbol{\mu}^{1},\theta,k_{1})-K_{N}(\mathbf{t},\delta,\boldsymbol{\mu}^{1},\theta,k_{1})\nonumber\\
&= (N_{l})^{1/2k_{1}}\exp(Nf-\langle\mathbf{t},\mathbf{N}^{1/2k_{1}}\boldsymbol{\mu}^{1}\rangle)
\nonumber\\
&\quad\times\int_{B(\boldsymbol{\mu}^{1},\delta_{1})\smallsetminus B(\boldsymbol{\mu}^{1},d)}\!\!\!\!\exp\Big(-\frac{N}{2}\langle\widetilde{\mathbf{J}}\mathbf{x},\mathbf{x}\rangle+\langle\mathbf{t},\mathbf{N}^{1/2k_{1}}\mathbf{x}\rangle\Big)I_{N}^{c}(\mathbf{x},\boldsymbol{\mu}^{1},\theta)d\mathbf{x}.
\end{align}
By (\ref{tortainarrivo}) we have:
\begin{equation}
I_{N}^{c}(\mathbf{x},\boldsymbol{\mu}^{1},\theta)\leq\exp(N\Phi(\mathbf{x})).
\end{equation}
By definition of the function $\Phi$ we get
\begin{align}
K_{N}(\mathbf{t},\delta_{1},\boldsymbol{\mu}^{1},\theta,k_{1})-K_{N}(\mathbf{t},\delta,\boldsymbol{\mu}^{1},\theta,k_{1})
&\leq (N_{l})^{1/2k_{1}}e^{Nf-\langle\mathbf{t},\mathbf{N}^{1/2k_{1}}\boldsymbol{\mu}^{1}\rangle}\nonumber\\
&\quad\times\int_{B(\boldsymbol{\mu}^{1},\delta_{1})\smallsetminus B(\boldsymbol{\mu}^{1},d)}\!\!\!\!\exp(-NG(\mathbf{x})+\langle\mathbf{t},\mathbf{N}^{1/2k_{1}}\mathbf{x}\rangle)d\mathbf{x}.
\end{align}
Making the change of variable
\begin{equation}
u_{l}=\mu_{l}+\dfrac{x_{l}}{N_{l}^{1/2k_{1}}}\quad l=1,\dots,n
\end{equation}
we obtain
\begin{equation}
K_{N}(\mathbf{t},\delta_{1},\boldsymbol{\mu}^{1},\theta,k_{1})-K_{N}(\mathbf{t},\delta,\boldsymbol{\mu}^{1},\theta,k_{1})\\\leq\int_{E}\exp\bigg(-N\bigg(B\bigg(\frac{\mathbf{u}}{\boldsymbol{\mu}^{1}+\mathbf{N}^{1/2k_{1}}};\boldsymbol{\mu}^{1}\bigg)\bigg)+\langle\mathbf{t},\mathbf{u}\rangle\bigg)d\mathbf{u}
\end{equation}
where 
\begin{equation}
E=\Big\{\Big|\Big|\frac{\mathbf{u}}{\mathbf{N}^{1/2k_{1}}}\Big|\Big|<\delta_{1}\Big\}\setminus\Big\{\Big|\Big|\frac{\mathbf{u}}{\mathbf{N}^{1/2k_{1}}}\Big|\Big|<\delta\Big\}.
\end{equation}
Observing that
\begin{equation}
E\subset\Big\{\Big|\Big|\frac{\mathbf{u}}{\mathbf{N}^{1/2k_{1}}}\Big|\Big|<\delta\Big\}^{c}
\end{equation}
it follows that as $N\rightarrow\infty$, for some $\epsilon_{0}>0$
\begin{equation}\label{stima2}
 K_{N}(\mathbf{t},\delta_{1},\boldsymbol{\mu}^{1},\theta,k_{1})-K_{N}(\mathbf{t},\delta,\boldsymbol{\mu}^{1},\theta,k_{1})=O(e^{N\epsilon_{0}})
\end{equation}
The statement (\ref{tesi.lemma.funzione.k}) follows by (\ref{stima1}) and (\ref{stima2}).\vspace{1cm}\\

{\bf Acknowledgments}: we thank an anonymous referee for pointing out to us an important reference.
We thank moreover Prof. C. Giardin\`a, Prof C. Giberti and F. Unguendoli for interesting discussions.


\end{document}